\documentclass{aa}

\usepackage{epsf}
\usepackage{graphics}

\begin{document}

\def\clipfig#1{\def\lbracket{[}\def\testit{#1}%
    \ifx\testit\lbracket\let\next=\optclipfig\else\let\next=\stdclipfig\fi%
    \next{#1}}
%
\newcommand {\hclipfig} [7] {\clipfig[#7]{#1}{#2}{#3}{#4}{#5}{#6}}
%
\def\usemodepsfig {\global\def\cfmode{x}\typeout{*** set clipfig to PSFIG mode ***}}
\def\usemodeepsf  {\global\def\cfmode{}\typeout{*** set clipfig to EPSF mode ***}}
\def\useunitmm    {\global\def\cfunit{x}\typeout{*** set clipfig to use mm as unit ***}}
\def\useunitcm    {\global\def\cfunit{}\typeout{*** set clipfig to use cm as unit ***}}
\def\clipfigsettings {\ifx\cfmode\empty\def\ccfmode{EPSF }\else\def\ccfmode{PSFIG }\fi%
    \ifx\cfunit\empty\def\ccfunit{cm }\else\def\ccfunit{mm }\fi%
    \typeout{*** current clipfig settings: \ccfmode mode, using \ccfunit as unit ***}}
%
%
%
%
\def\stdclipfig#1#2#3#4#5#6{\ifx\cfmode\empty%
    \let\next=\eclipfig\else\let\next=\pclipfig\fi%
    \next{#1}{#2}{#3}{#4}{#5}{#6}}
\def\optclipfig#1#2]#3#4#5#6#7#8{\ifx\cfmode\empty%
    \let\next=\ehclipfig\else\let\next=\phclipfig\fi%
    \next{#3}{#4}{#5}{#6}{#7}{#8}{#2}}
%
%
%
\newcommand {\pclipfig}[6] {\ifx\cfunit\empty%
        \psfig{figure=#1.ps,width=#2cm,bbllx=#3cm,bblly=#4cm,bburx=#5cm,%
           bbury=#6cm,clip=}\else%
        \psfig{figure=#1.ps,width=#2mm,bbllx=#3mm,bblly=#4mm,bburx=#5mm,%
           bbury=#6mm,clip=}\fi}
\newcommand {\phclipfig}[7] {\ifx\cfunit\empty%
        \hspace{#7cm}\psfig{figure=#1.ps,width=#2cm,bbllx=#3cm,bblly=#4cm,%
           bburx=#5cm,bbury=#6cm,clip=}\else%
        \hspace{#7mm}\psfig{figure=#1.ps,width=#2mm,bbllx=#3mm,bblly=#4mm,%
           bburx=#5mm,bbury=#6mm,clip=}\fi}
%
%
%
\newcommand {\eclipfig}[6]{%
  \ifx\cfunit\empty\epsfxsize=#2cm\else\epsfxsize=#2mm\fi%
  \epsfclipon\epsfverbosetrue%
  \cfcmtopspts{#3}\cfllxi=\cftempi\cfllxf=\cftempf%
  \cfcmtopspts{#4}\cfllyi=\cftempi\cfllyf=\cftempf%
  \cfcmtopspts{#5}\cfurxi=\cftempi\cfurxf=\cftempf%
  \cfcmtopspts{#6}\cfuryi=\cftempi\cfuryf=\cftempf%
  \def\cfstra{\number\cfllxi.\number\cfllxf}%
  \def\cfstrb{\number\cfllyi.\number\cfllyf}%
  \def\cfstrc{\number\cfurxi.\number\cfurxf}%
  \def\cfstrd{\number\cfuryi.\number\cfuryf}%
  \hbox{\epsfbox[{\cfstra} {\cfstrb} {\cfstrc} {\cfstrd}]{#1.ps}}}
\newcommand {\ehclipfig}[7]{%
  \ifx\cfunit\empty\epsfxsize=#2cm\else\epsfxsize=#2mm\fi%
  \epsfclipon\epsfverbosetrue%
  \cfcmtopspts{#3}\cfllxi=\cftempi\cfllxf=\cftempf%
  \cfcmtopspts{#4}\cfllyi=\cftempi\cfllyf=\cftempf%
  \cfcmtopspts{#5}\cfurxi=\cftempi\cfurxf=\cftempf%
  \cfcmtopspts{#6}\cfuryi=\cftempi\cfuryf=\cftempf%
  \def\cfstra{\number\cfllxi.\number\cfllxf}%
  \def\cfstrb{\number\cfllyi.\number\cfllyf}%
  \def\cfstrc{\number\cfurxi.\number\cfurxf}%
  \def\cfstrd{\number\cfuryi.\number\cfuryf}%
  \ifx\cfunit\empty\hspace{#7cm}\else\hspace{#7mm}\fi%
  \hbox{\epsfbox[{\cfstra} {\cfstrb} {\cfstrc} {\cfstrd}]{#1.ps}}%
  \vspace{-1mm}}
%
%
%
\newdimen\cfllxi \newdimen\cfllyi  \newdimen\cfurxi  \newdimen\cfuryi
\newdimen\cfllxf \newdimen\cfllyf  \newdimen\cfurxf  \newdimen\cfuryf
\newdimen\cftemp \newdimen\cftempi \newdimen\cftempf
\newdimen\cfpspoint \cfpspoint=1bp
%
%
%
\newcommand{\cfcmtopspts}[1]{\ifx\cfunit\empty%
  \cftemp=#1cm\else\cftemp=#1mm\fi%
  \multiply\cftemp10\divide\cftemp\cfpspoint%
  \cftempf=\cftemp\divide\cftemp10\cftempi=\cftemp\multiply\cftemp10%
  \advance\cftempf-\cftemp}
%
%
\def\cfmode{}\def\cfunit{}\clipfigsettings
%

\useunitmm


\newcommand{\lb}{$\lambda$}
\newcommand{\sm}[1]{\footnotesize {#1}}
\newcommand{\inft}{$\infty$}
\newcommand{\vlv}{$\nu L_{\rm V}$}
\newcommand{\lv}{$L_{\rm V}$}
\newcommand{\lx}{$L_{\rm x}$}
\newcommand{\lsoft}{$L_{\rm 250eV}$}
\newcommand{\lhard}{$L_{\rm 1keV}$}
\newcommand{\vlsoft}{$\nu L_{\rm 250eV}$}
\newcommand{\vlhard}{$\nu L_{\rm 1keV}$}
\newcommand{\vlir}{$\nu L_{60\mu}$}
\newcommand{\ax}{$\alpha_{\rm x}$}
\newcommand{\aopt}{$\alpha_{\rm opt}$}
\newcommand{\aox}{$\alpha_{\rm ox}$}
\newcommand{\aoxh}{$\alpha_{\rm oxh}$}
\newcommand{\airhard}{$\alpha_{\rm 60\mu-hard}$}
\newcommand{\aoxsoft}{$\alpha_{\rm ox-soft}$}
\newcommand{\aio}{$\alpha_{\rm io}$}
\newcommand{\aixs}{$\alpha_{\rm ixs}$}
\newcommand{\aixh}{$\alpha_{\rm ixh}$}
\newcommand{\hb}{H$\beta$}
\newcommand{\nh}{$N_{\rm H}$}
\newcommand{\nhgal}{$N_{\rm H,gal}$}
\newcommand{\nhfit}{$N_{\rm H,fit}$}
\newcommand{\ale}{$\alpha_{\rm E}$}
\newcommand{\cts}{$\rm {cts\,s}^{-1}$}
\newcommand{\pl}{$\pm$}
\newcommand{\kev}{\rm keV}
\newcommand{\rb}[1]{\raisebox{1.5ex}[-1.5ex]{#1}}
\newcommand{\ten}[2]{#1\cdot 10^{#2}}
\newcommand{\msun}{$M_{\odot}$}
\newcommand{\dM}{$\dot{M}$}
\newcommand{\dMM}{$\dot{M}/M$}
\newcommand{\dMedd}{\dot M_{\rm Edd}}
\newcommand{\kms}{km\,$\rm s^{-1}$}
\newcommand{\mum}{$\mu$m}
\newcommand{\pa}{Pa$\alpha$}
\newcommand{\brg}{Br$\gamma$}
\newcommand{\brd}{Br$\delta$}
\newcommand{\ha}{H$\alpha$}


\title{Near Infrared observations of Soft X-ray selected 
AGN\thanks{Based on observations performed at the 2.7m telecope of the
McDonald Observatory, Texas}
}
\author{D. Grupe\inst{1},
\and H.-C. Thomas\inst{2}
}
\offprints{\\ D. Grupe (dgrupe@xray.mpe.mpg.de)}
\institute{MPI f\"ur extraterrestrische Physik, Postfach 1312, 
85741 Garching, Germany
\and MPI f\"ur Astrophysik, Karl-Schwarzschild-Str. 1, 85748 Garching, Germany
}
\date{received 19. December 2001; accepted 20. February 2002}

\abstract{
We report the results of near infrared observations of 19 soft X-ray
selected AGN. The goal of the observations was 
to search for 
strong, narrow \pa~ or \brg~ emission lines, as a
sign of nuclear starbursts.  We found
Pa$\alpha$~ emission in the spectra of 11 sources and Br$\gamma$~ in at
least five. Strong NIR emission has been found in two sources, CBS 126 and Mkn
766, both objects with strong [OIII]$\lambda$5007 emission, weak Fe\,II emission and
wavelength dependent degree of polarization in the optical. Classical
Narrow Line Seyfert 1 galaxies do not show exceptionally strong NIR emission
lines. We present the results of our study and discuss how our findings fit into
an evolutionary scheme of AGN.
}

\maketitle

\section{Introduction}

Optical spectra of AGN exhibit many significant relationships.
For example,
increasing strength of Fe\,II emission from the broad line region (BLR) 
corresponds to decreasing width of
H$\beta$\ from the BLR, and decreasing strength of
[O\,III]\,$\lambda$5007 emission from the narrow line region (NLR)
(Boroson \& Green 1992, Grupe et al. 1999).
ROSAT (Tr\"umper 1983), 
with its high sensitivity to soft X-rays, has played an important
role in showing that these trends correspond to an increasing steepness of the
soft X-ray spectra (Boller et al. 1996,  Wang et al. 1996,
Laor et al. 1997, Grupe et al. 1998a).
Correlation analyses show that the dominant spectrum-to-spectrum variation
in AGN samples can be reduced to a linear combination of
these emission line and continuum parameters -- the `First Principal 
Component' (Boroson \& Green 1992).     

The `First Principal Component' (or Eigenvector 1) almost certainly represents
the Eddington ratio L/L$_{\rm Edd}$, where L is the bolometric luminosity, and
L$_{\rm Edd}$ is the upper limit to a Black Hole's luminosity for which radiation
forces on the accreting matter balance gravitational forces (e.g., Boroson 2002).
Narrow Line Seyfert 1 nuclei (NLS1), with their steeper X-ray spectra, 
stronger Fe\,II, weaker [O\,III] emission and narrower
\hb~ represent higher Eddington ratios, and are therefore suggested to be an earlier
stage of activity (Grupe 1996, Grupe et al. 1999, Mathur 2000), or a rejuvenated phase
of Seyfert 1 activity. 
The association between AGN activity and starbursts is well documented 
(e.g. Gonz\'ales Delgado 2001, 
Cid Fernandes et al. 2001 and references therein).
Starbursts, perhaps triggered by galaxy mergers or interactions, are usually associated
with fueling  the central engine, although it has been suggested that AGN activity
triggers the starbursts. The AGN-starburst connection has been demonstrated for
Seyfert 2 galaxies where about 30-50\% do show massive young stars within
$\sim$ 300 pc of the center (Cid Fernandes et al. 2001).
Consistent with this, Wills et al. (1999) suggested that starbursts could provide the
high-column densities of dense, Fe-enriched gas suggested by Principal Component 1
spectra.  Starburst galaxies are known to be bright in the infrared. 
Therefore
we would expect to see higher IR luminosities in NLS1 than in broad line Seyfert
1s (BLS1).  We found that soft X-ray selected AGN on the average
 show slightly higher IR luminosities than AGN from hard X-ray selected
samples (Grupe et al. 1998a, Moran et al. 1996). 
However, among the most luminous infrared
objects, the ultra-luminous IRAS galaxies (ULIRGs), AGN appear to be rare
(Murphy et al. 2001) and recently Ivanov et al. (2000) found from a K-band
spectroscopy study of 33 Seyfert galaxies that there is no evidence for strong
starbursts among their sources. 
At optical/NIR wavelengths one way to indicate a
starburst in nearby bright galaxies is to measure the strength of the CaII
triplett ($\lambda\lambda$8498, 8542, 8662 \AA; e.g. Terlevich et al. 1990).
However, 
while in the optical it is not (easily) possible to distinguish 
between the activity from the nucleus itself and a circumnuclear starburst,  
it is
(easier) in the infrared (e.g. Genzel \& Cesarsky 2000, Laurent et al. 2000).

In principle, 
 near infrared (NIR) 
spectroscopy provides a tool to distinguish between young and old
starbursts by searching for strong hydrogen emission lines
from young stars or CO
absorption lines at 2.3\mum~ from old stars (Heisler \& De Robertis 1999).  
These CO absorption lines have been found to be common in obcured AGN while
being absent in Seyfert 1s (Oliva et al. 1999).

We observed 19 sources of our complete
sample of 113 soft X-ray
selected AGN (Grupe et al. 2001) by NIR spectroscopy in the K-band. The sources
were selected from the ROSAT All-Sky Survey (RASS, Voges et al. 1999) by
their count rate and hardness ratio (Thomas et al. 1998, Grupe et al.
1998a). In this paper we present the results of a first observing 
run to study
the soft X-ray selected AGN of our sample in the infrared. We have organized
this paper as follows: In Sect. \ref{observe} we describe the observation and the
data reduction and analysis, in Sect. \ref{res-sect} the results of our NIR
study are presented, and are discussed in Sect. \ref{discuss}. 

\section{\label{observe} Observations and data reduction}
The near-infrared spectroscopy of our AGN subsample used the
CoolSpec spectrograph (Lester et al. 2000)
on the 2.7m telescope at McDonald Observatory in Texas. We used grating
\#3310-FL-906 in the K-band in first order with a resolution
$\lambda/\Delta\lambda$=268 and a 256$\times$256 NICMOS3 CCD 
($\Delta\lambda$ at 2.2\mum $\equiv$ 2.6 pixel). 
Table \ref{obslist} lists all sources with the
optical positions, redshifts, K magnitudes from the 2 Micron All Sky Survey
(2MASS) second incremental release catalogue, observed K magnitude K$_{\rm obs}$,
and observing dates and times.  During the whole run the weather
conditions were rather poor,  
therefore absolute flux calibrations are not very accurate.
In order to obtain these we 
observed  for each object a nearby bright star of spectral type A to
G. The spectral types were taken from the Bright Source Catalogue, 5th revised
edition (Hoffleit \& Jaschek 1991).
For some of our comparison stars the K magnitudes were known, for
others the K magnitudes were derived from the B and V magnitudes using the
relation V-K = (2.29\pl0.01)$\times$(B-V). This relation was obtained by 
fitting the
colours in a sample of 196 infrared standard stars common to the list of 
Van der Bliek et al. (1996) and the Bright Star Catalogue, 5th Revised Ed.
(Hoffleit \& Jaschek 1991). These K magnitudes were converted into flux
densities at 2.2 \mum~ using the relation
K = 14.0 - 2.5 log($F_{2.2\mu}$) (Van der Bliek et al. 1996) 
with the flux density in units of 10$^{-15}$ W m$^{-2}$ \mum$^{-1}$.
The flux density
distribution in our observing window was then computed from the flux density at
2.2 \mum~ 
assuming a
Rayleigh-Jeans tail for the spectral shape.

The flux-calibrated spectrum of the AGN can then be calculated from the ratio of
the observed spectra of AGN and comparison star, 
taking into account the different exposure times.
The flux densities at 2.2\mum~ thus obtained for our AGN were converted into K
magnitudes using the relation given above. 
In Table \ref{obslist} we list these K magnitudes together with 
those given in the 2MASS catalogue. 
In most cases the observed magnitude is
less than the one given in the 2MASS, because the poor weather conditions are
more likely to reduce the flux for the long exposures of the objects than for
the short exposures of the comparison stars. 
In this way we can scale the observed AGN
spectrum, to account for variations in the
extinction between source and comparison star and we
are able to get more reliable absolute infrared
fluxes that can be compared to our
measurements in the optical range. As listed in Table \ref{obslist} for six
sources K magnitudes were not yet available in the 2MASS. Therefore we give
line fluxes in Table \ref{results} only for those sources for which we 
could do the scaling as described above.  
         
Wavelength calibrations were obtained by fitting a linear relation to the line
positions of Argon, Neon, and Xenon. Sky subtraction was done by nodding the
telescope between two positions in a star-sky-sky-star sequence. For our targets
such a sequence took 8 min and was repeated several times, while for the bright
comparison stars a much shorter exposure time of 4 to 20 s per sequence was
sufficient.

\begin{table*}
\caption{\label{obslist} List of the AGN; $\alpha_{2000}$ and $\delta_{2000}$
are the optical positions (Grupe et al., 2001), K magnitudes are taken from the
2MASS catalogue and from the observed spectrum (before scaling), 
and T$_{obs}$ lists the total observing time per source in
minutes
}
\begin{flushleft}
\begin{tabular}{rlccccccrcc}
\hline\noalign{\smallskip}
\# & Name & $\alpha_{2000}$ & $\delta_{2000}$ & z & K$_{\rm 2MASS}$ 
& K$_{\rm obs}$ & Obs date & T$_{obs} $ & comp. star & spec. type \\
\noalign{\smallskip}\hline\noalign{\smallskip}
1 & RX J0859.0+4846 & 08 59 02.9 & +48 46 09 & 0.083 & --- & 12.3 &
2001-02-18 & 24 & HR 2692 & G9V \\
2 & Mkn 110 & 09 25 13.0 & +52 17 12 & 0.035 &  --- & 12.5 &
2001-02-18 & 32 & HR 4096 & A2V  \\
3 & PG0953+414 & 09 56 52.4 & +41 15 22 & 0.234 & 12.5 & 12.4 &
2001-02-19 & 16 & HR 3811 & F2V \\
4 & RX J1005.7+4332 & 10 05 41.9 & +43 32 41 & 0.178 & 12.7 & 12.9 &
2001-02-17 & 32 & HR 4067 & F7V \\
5 & RX J1007.1+2203 & 10 07 10.2 & +22 03 02 & 0.083 & 14.2 & 14.0 &
2001-02-19 & 32 & HR 4012 & F9V \\
6 & CBS 126 & 10 13 03.2 & +35 51 24 & 0.079 & 12.6 & 13.0 &
2001-02-18 & 24 & HR 4096 & A2V \\
7 & RX J1034.6+3938 & 10 34 38.6 & +39 38 28 & 0.044 & 12.7 & 13.8 &
2001-02-17 & 40 & HR 4067 & F7V \\
8 & RX J1117.1+6522 & 11 17 10.1 & +65 22 07 & 0.147 & 13.0 & 14.0 &
2001-02-19 & 32 & HR 3391 & G1V \\
9 & Ton 1388 & 11 19 08.7 & +21 19 18 & 0.177 & 11.5 & 12.2 &
2001-02-19 & 16 & HR 4012 & F9V \\
10 & PG 1211+143 & 12 14 17.7 & +14 03 13 & 0.082 & --- & 12.0 &
2001-02-19 & 24 & HR 4864 & G7V \\
11 & Mkn 766 & 12 18 26.6 & +29 48 46 & 0.013 & 10.6 & 10.6 &
2001-02-18 & 16 & HR 4096 & A2V \\
12 & IC 3599 & 12 37 41.2 & +26 42 28 & 0.021 & 13.5 & 13.8 &
2001-02-18 & 32 & HR 4096 & A2V \\
13 & IRAS 12397+3333 & 12 42 10.6 & +33 17 03 & 0.044 & --- & 12.0 &
2001-02-18 & 24 & HR 4845 & G0V \\
14 & RX J1304.2+0205 & 13 04 17.0 & +02 05 37 & 0.229 &  --- & 15.6 &
2001-02-19 & 16 & HR 4864 & G7V \\
15 & PG 1307+085 & 13 09 47.0 & +08 19 48 & 0.155 & --- &  13.1 &
2001-02-19 & 24 & HR 4864 & G7V \\
16 & RX J1355.2+5612 & 13 55 16.6 & +56 12 45 & 0.122 & 13.3 & 14.1 &
2001-02-19 & 24 & HR 5280 & A2V \\
17 & PG 1402+261 & 14 05 16.2 & +25 55 34 & 0.164 & 12.2 & 12.8 &
2001-02-18 & 32 & HR 4845 & G0V \\
18 & Mkn 478 & 14 42 07.5 & +35 26 23 & 0.077 & 11.1 & 12.2 &
2001-02-18 & 32 & HR 5569 & A2V  \\
19 & Mkn 493 & 15 59 09.7 & +35 01 48 & 0.032 & 11.8 & 12.0 &
2001-02-19 & 16 & HR 5280 & A2V \\
\noalign{\smallskip}\hline\noalign{\smallskip}
\end{tabular}
\end{flushleft}

\end{table*}

\section{\label{res-sect} Results}

Figure \ref{nir-spec} displays all NIR spectra of the 19 soft X-ray selected AGN
we observed as well as a mean spectrum of the atmospheric absorption.
Out of 12 sources with redshifts that put Pa$\alpha$ into the observable
window  the Pa$\alpha$ line was clearly detected in 11 cases. The
much weaker Br$\gamma$ line should have been visible for 11 sources.
Of those sources it was clearly detected in Mkn 766 and possibly in RX J0859+48,
Mkn 110, CBS 126, and IRAS12397+3333. In the other 6 sources, where
Br$\gamma$ is marked in the spectrum, 
it is not clear if this is noise or a real
line. Therefore in Table \ref{results} the Br$\gamma$ measurements 
of those sources are noted as upper limits. Upper limits were estimated from 
the noise peaks in the spectrum.
Due to the redshift of the sources for only a few objects it is possible to get
the Pa$\alpha$ as well as the Br$\gamma$ line.

Table \ref{results} lists the results from the line measurements. 
 The table contains the object name (as
given in Table \ref{obslist}), the Seyfert type, the rest frame equivalent
widths of Pa$\alpha$ and Br$\gamma$ and the rest frame FWHM(Pa$\alpha$). The
FWHM(Pa$\alpha$) (in \AA)
was deconvolved by the spectral resolution of about 180 \AA,
using $\rm FWHM(Pa\alpha_{corr})^2~=~FWHM(Pa\alpha_{observed})^2~-~180^2$.
The Pa$\alpha$ line widths are in agreement with other studies in the literature
(e.g. Hines, 1991). The table also contains the line
fluxes of Pa$\alpha$ and Br$\gamma$ and the Pa$\alpha$/H$\beta$ line ratio. 
The H$\beta$ line flux was measured from FeII subtracted optical
spectra (Grupe et al. 1999). The cases are marked in the table,
for which absolute fluxes can not be given, because no 2MASS data are available
yet.

Two of the sources listed in Table \ref{results} show exceptionally high 
 Pa$\alpha$/H$\beta$ flux ratios, CBS 126 and RX J1355.2+5612. From the case B
 scenario (Hummer \& Storey 1987)
 a Pa$\alpha$/H$\beta$ ratio=0.332 is expected.  For only one AGN, Mkn 766 
 a flux ratio Br$\gamma$/H$\beta$ can be measured. For this AGN the
Br$\gamma$/H$\beta$
 flux ratio is about 3 times higher than expected for a case B (0.0275,
 Hummer \& Storey, 1987).

\begin{table*}
\caption{\label{results} Results of the analysis of the NIR data. The 
rest-frame equivalent widths EW
are given in \AA, the FWHM(Pa$\alpha$) in \kms, and the 
fluxes for Pa$\alpha$ and Br$\gamma$ in units of 10$^{-17}$ W m$^{-2}$ and 
10$^{-18}$ W m$^{-2}$, respectively. For comparison the FWHM(H$\beta$) is listed
in \kms.
}
\begin{flushleft}
\begin{tabular}{rlccrrrrcr}
\hline\noalign{\smallskip}
& & & \multicolumn{2}{c}{EW} &  FWHM & \\
\rb{\#} & \rb{Name}  & \rb{Type} & Pa $\alpha$ & Br $\gamma$  & Pa$\alpha$ &
\rb{$F_{\rm Pa \alpha}$} & \rb{$F_{\rm Br \gamma}$}  & 
\rb{F(Pa$\alpha$)/F(H$\beta$)} & \rb{FWHM(H$\beta$)} \\
\noalign{\smallskip}\hline\noalign{\smallskip}
1 & RX J0859.0+4846 & S1 & 160\pl25 & 10\pl5  & 3000\pl100 & 7.5\pl0.5$^2$ & 
3.5\pl1.0$^2$  &  0.40$^{2}$ & 2900$^{~}$ \\
2 & Mkn 110 & NLS1 & ---$^{3}$ & 10\pl5  & --- & --- & 3.7\pl0.7$^2$ &   --- &
1500$^{~}$ \\
3 & PG0953+414 & S1 & 150\pl35 & ---$^{3}$  & 3000\pl500 & 6.0\pl0.5 & ---$^{~}$ &  
--- & 3100$^4$ \\
4 & RX J1005.7+4332 & S1 & 50\pl30 & ---$^{3}$  & 1700\pl1200 & 1.1\pl0.4 & ---$^{~}$ 
&  0.50 & 2990$^{~}$ \\
5 & RX J1007.1+2203 & NLS1  & 110\pl30 & 25$^1$ & 3100\pl450 & 0.9\pl0.2 & 0.2$^1$ 
& 0.22 & 1400$^{~}$ \\
6 & CBS 126 & S1 & 310\pl20 & 5\pl3 & 1600\pl200 & 12.0\pl0.5 & 0.3\pl0.1  & 1.45 
& 2850$^{~}$ \\
7 & RX J1034.6+3938 & NLS1 & ---$^{3}$ & 10$^1$ & --- &  --- & 0.5$^1$ &   
--- & 750$^{~}$ \\
8 & RX J1117.1+6522 & NLS1 & 120\pl30 & ---$^{3}$  & 3200\pl550 & 2.5\pl0.5 & 
---$^{~}$ & 0.60 & 2160$^{~}$ \\
9 & Ton 1388 & S1 & 85\pl10 & ---$^{3}$ &  2400\pl400 & 8.0\pl1.0 & ---$^{~}$ & 0.20 
& 2920$^{~}$ \\
10 & PG 1211+143 & NLS1 & 200\pl20 & 20$^1$ & 1700\pl100 & 12.0\pl0.4$^2$   & 
0.3$^1$  & 0.34$^{2}$ & 1900$^{~}$ \\
11 & Mkn 766 & NLS1  & ---$^{3}$ & 6\pl2 & --- &  --- & 1.4\pl0.3$^{~}$  & 
--- & 1360$^{~}$ \\
12 & IC 3599 & S2 & ---$^{3}$ & 10$^{1}$ & --- & --- & 4.0$^1$ & --- & 635$^{~}$
\\
13 & IRAS 12397+3333 & NLS1 & ---$^{3}$ & 10\pl5 & --- & --- & 5.0\pl2.0$^2$  & 
--- & 1900$^{~}$ \\
14 & RX J1304.2+0205 & NLS1 & ---$^{~}$ & ---$^{3}$ & ---  &  --- & ---$^{~}$ & 
--- & 1400$^5$ \\
15 & PG 1307+085  & S1 & 160\pl30 & ---$^{3}$ & 2200\pl100 & 2.7\pl0.5$^2$ & 
---$^{~}$ & 0.15$^{2}$ & 4000$^{~}$ \\
16 & RX J1355.2+5612 & NLS1 & 50\pl30 & ---$^{3}$ & 1600\pl1000 & 1.2\pl0.6 & ---$^{~}$ 
& 3.50 & 1780$^{~}$ \\
17 & PG 1402+261 & NLS1 & 65\pl15 & ---$^{3}$ & 2300\pl250 & 3.7\pl0.3 & ---$^{~}$ & 
0.28 & 1700$^{~}$ \\
18 & Mkn 478 & NLS1 & 50\pl2 & 2$^{1}$ & 1550\pl200 &  8.0\pl0.5 & 4.0$^1$ & 0.27 
& 1915$^{~}$ \\
19 & Mkn 493 & NLS1 & ---$^{3}$ & 10$^1$ & --- & --- & 8.0$^1$ & --- & 900$^{~}$\\
\noalign{\smallskip}\hline\noalign{\smallskip}
\end{tabular}
\end{flushleft}

$^1$ Upper limit \\
$^2$ No 2MASS data available, and therefore no scaling possible to get absolute
fluxes. \\
$^3$ Out of the K-band window \\
$^4$ Boroson \& Green, 1992 \\
$^5$ FWHM(H$\alpha$)

\end{table*}

\def \charthoffset {\hspace{0.2cm}} \def \charthsep {\hspace{0.3cm}}
\def \chartvsepcap {\vspace{0.3cm}}
\def \chartvsep {\vspace{0.1cm}}
\newcommand{\putchart}[1]{\clipfig{./#1}{58}{45}{153}{145}{194}
}
\newcommand{\chartline}[3]{\parbox[t]{18cm}{
 
\noindent\charthoffset\putchart{#1}\charthsep\putchart{#2}\chartvsep\putchart{#3}\chartvsep}}

\rule{0mm}{0mm}

\begin{figure*}
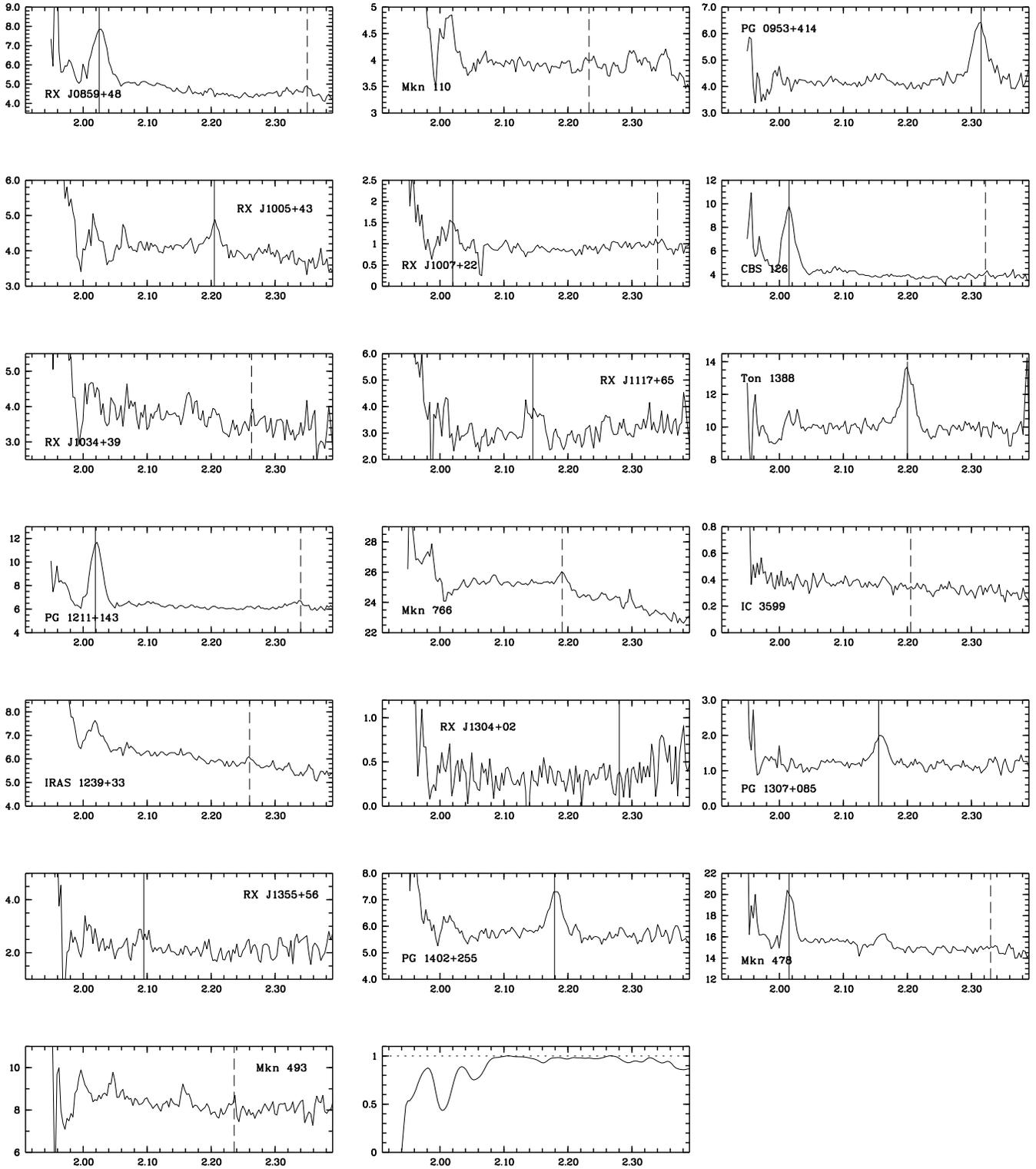

\caption[ ]{\label{nir-spec}
NIR spectra of the 19 observed AGN;
Fluxes are given in units of 10$^{-15}$ W m$^{-2}$ $\rm \mu m^{-1}$ and
the wavelength in $\mu$m. The solid line displays the position of the Pa$\alpha$
line and the dashed line the position of Br$\gamma$.
The last spectrum shows the atmospheric transmission
normalized to unity.
}
\chartvsepcap
\chartline{H3373_F1}{H3373_F2}{H3373_F3}
\chartline{H3373_F4}{H3373_F5}{H3373_F6}
\chartline{H3373_F7}{H3373_F8}{H3373_F9}
\chartline{H3373_F10}{H3373_F11}{H3373_F12}
\chartline{H3373_F13}{H3373_F14}{H3373_F15}
\chartline{H3373_F16}{H3373_F17}{H3373_F18}
\chartline{H3373_F19}{H3373_F20}{test-spec}
\end{figure*}

\section{\label{discuss} Discussion and conclusions}

The main motivation for our study of NIR spectra of soft X-ray selected AGN was
to search for signatures of a young starburst. The idea is that NLS1 are young
AGN and therefore should show a young starburst which should be visible
in the NIR spectrum. Signs of this young starburst would be the detection of
strong Pa$\alpha$ and/or Br$\gamma$ emission lines in excess of what is expected
from the broad H$\beta$ and H$\alpha$ line emission. 
 
 CBS 126 and RX J1355.2+5612 are the only two sources which show very high
 Pa$\alpha$/H$\beta$ line rations.
 Both sources have an H$\alpha$/H$\beta$ ratio = 3.5 (Grupe et al.
 1999). The internal reddening of the two sources 
 is close to the average found for the whole sample (Grupe et
 al.1998b). Also the spectral analysis of the RASS data shows that the
 cold (neutral) absorption in soft X-rays is in agreement with no or minor
 intrinsic absorption. For RX J1355.2+5612 the signal-to-noise ratio at the
 Pa$\alpha$ line is very low and  therefore the errors in the flux measurements
 are large. This explains the high ratio in this source. The situation is
 different for CBS 126. Even though the line might be slightly affected by the
 correction of the atmospheric
 absorption troughs at around 2\mum, the line is still strong
 and the high flux ratio Pa$\alpha$/H$\beta$ seems to be real. Displaying the 
 Pa$\alpha$/H$\beta$ ratio of CBS 126 in a H$\alpha$/H$\beta$ vs.
 Pa$\alpha$/H$\alpha$ diagram (e.g. Lacy et al., 1982, 
 Evans \& Natta 1989, Thompson 1992) shows
 that CBS 126 is far off most sources towards a high Pa$\alpha$/H$\alpha$ value.
 The other three sources for which the H$\alpha$/H$\beta$ ratio was available
 and  a scaling of the spectrum 
 by the 2MASS K magnitude was possible, RX J1007.1+2203, Ton 1388, and Mkn 478,
 are almost on the expected Pa$\alpha$/H$\beta$=0.28 line in Fig. 2
 in Lacy et al. (1982), which means they are not absorbed. Correcting
 CBS 126 for reddening reduces the Pa$\alpha$/H$\alpha$ to 0.41, which is still
 a strong excess in the Pa$\alpha$ line. 
 A possible explanation is that this is a
 contribution from a nuclear starburst. In order to verify this, observations on
 this object with higher resolution spectroscopy have to be performed.

 The clearest detection of a Br$\gamma$ line was found in Mkn 766. For this
 source one has to take into account that it does
 show intrinsic optical reddening that may explain part of the high
 Br$\gamma$/H$\beta$ flux ratio.
 The H$\alpha$/H$\beta$ ratio is 5.9 (Grupe
 et al. 1998b) and its optical spectral slope $\alpha_{\rm opt}$ = 1.9, which is
 high compared to the rest of the sample ($<$\aopt$>$=1.0; Grupe et al. 1998b).
 It is also a source that shows a relatively high degree of
 polarization of 2.3\% (Goodrich 1989), while most of the sources in our sample
 do not show significant polarization (Grupe et al. 1998b). Arguing also for a
 significant absorption in Mkn 766 is the finding of Lee et al. (2001) from
 X-ray spectra derived from CHANDRA's LETG of a dusty warm absorber in
 MCG--6-30-15 which has a similar X-ray spectrum as Mkn 766. However,  on the
 other hand, Branduardi-Raymont et al. (2001) argued from their X-ray
 spectroscopy studies with XMM-Newtons's Reflection Grating Spectrometer (RGS,
 den Herder et al. 2001) that the X-ray spectral features of Mkn 766 and
 MCG--6-30-15 can not be explained by a dusty warm absorber alone and proposed
 that they are caused by 
 gravitational redshift and relativistic broadening effects in
 the vicinity of a Kerr black hole. 

The question remains, why the classical NLS1 in our sample, 
such as Mkn 478 or Mkn
493, showing strong FeII emission and weak [OIII], not show exceptionally
strong NIR emission lines? The two sources that clearly have strong NIR emission
line, CBS 126 and Mkn 766, are not classical NLS1. Both have strong [OIII],
relatively weak FeII, and are both polarized in the optical. So far with
our current data set of 19 sources, we cannot give a final answer to the
question, simply for statistical reasons. Just recently, NIR spectra of two more
NLS1 of our soft X-ray sample have been published, Mkn 335 and Mkn 1044
(Rodr\'iguez-Ardila et al., 2002). Using their Pa$\alpha$ fluxes, we derived
Pa$\alpha$/H$\beta$ ratios of 0.29 and 0.11, respectively. 
Taking the starburst galaxy sample of Coziol et al. (2001) for estimating what
contribution we can expect from a usual starburst, the range of the equivalent
widths EW(Br$\gamma$) is between 2\AA-260 \AA. Based on these values we can exclude
a large starburst in any of AGN in our sample.
We have to extend the NIR sample in
order to derive
a more secure answer about which class has strong NIR emission
lines. This would bring us back to our original question
about the age of an AGN. Maybe the classical NLS1 are not the youngest AGN and
maybe that those AGN which show also stronger [OIII] would be younger. Two
facts argue for this scenario: 1) starburst galaxies always show strong
[OIII]/H$\beta$ ratio (e.g Osterbrock, 1989) 
and 2) recently Oshuga \& Umemura (2001) suggested
that Seyfert 2 galaxies are the younger AGN. In this case the strong starburst
in classical NLS1 is already passed. 
If this is true we may ask what happens then to the only Seyfert 2 galaxy in our
sample, IC 3599\footnote{Please note: IC 3599 can be classified as a Seyfert 1.9
galaxies based on higher resolution spectroscopy; Komossa \& Bade, 1999)}.
Shouldn't we see also strong NIR emission lines in the spectrum
of this AGN? The answer is in principle yes, but the H$\beta$ flux in this
source is already rather low (see e.g. Grupe et al. 1995) and assuming a case B
ratio of H$\beta$/Br$\gamma$=0.0275 (Hummer \& Storey 1987) we end up with the
upper limit we measured in this object. In other words, IC 3599 shows a line
ratio as expected, but not stronger than this either. It is interesting to note
that IC 3599 has shown a strong X-ray outburst followed by a response in its
optical emission lines (Grupe et al. 1995; Brandt et al. 1995). We could have
expected that when the light front passes 
through the center of the AGN that we would also see strong
NIR lines in its spectrum. However, it seems that the light passed already the
NIR emitting region. It would be interesting for future discoveries of X-ray
transient AGN or galaxies to follow them also in the infrared in order to
localize the NIR line emitting regions.

Our study of NIR spectra has shown that NIR spectroscopy of soft X-ray selected
AGN gives interesting results. 
In 11 cases we were able to detect Pa$\alpha$ emission.
However, a future task should be to
complete the NIR observations for the
whole sample to give a more secure statement of the development of
AGN. Performing the spectroscopy with higher resolution would help to
search for line asymmetries as they have been reported for H$\beta$ emission
(e.g. Boroson \& Green, 1992). However, often strong
H$\beta$ asymmetries have been especially reported for those
NLS1 with very strong FeII emission. Therefore the results of those H$\beta$ line
measurements might be influenced by the
contamination of the H$\beta$ line by FeII blends. The Pa$\alpha$
line is not affected by FeII contamination and therefore a possible line
asymmetry can be studied more accurately than in the \hb~line.  

\acknowledgements{ 
We want to thank Dr. Bev Wills for her help with the observing run and her
comments on this paper.
We also want to thank Dr. Dan Lester for his help and support during the
CoolSpec observing run. 
Many thanks also to Drs. Mario Gliozzi, Stefanie
Komossa, and Wolfgang Brinkmann
for their comments and suggestions on the manuscript.
This research has made use of the NASA/IPAC Extragalactic
Database (NED) which is operated by the Jet Propulsion Laboratory, Caltech,
under contract with the National Aeronautics and Space Administration. 
The ROSAT project is supported by the Bundesministerium f\"ur Bildung
und  Forschung (BMBF/DLR) and the Max-Planck-Society.

This paper can be retrieved via WWW: \\
http://www.xray.mpe.mpg.de/$\sim$dgrupe/research/refereed.html}

   \end{document}